# A Novel Magnetic Material by Design: Observation of $Yb^{3+}$ with Spin-1/2 and Possible Superconducting Trace in $Yb_xPt_5P$

*Xin Gui,[1] Tay-Rong Chang[2,3] Kaya Wei,[4] Marcus J. Daum,[5] David E. Graf,[4] Ryan. E. Baumbach,[4,6] Martin Mourigal,[5*] Weiwei Xie[1*]*

[1] Department of Chemistry, Louisiana State University, Baton Rouge, LA, USA 70803

[2] Department of Physics, National Cheng Kung University, Tainan, Taiwan 70101

[3] Center for Quantum Frontiers of Research & Technology (QFort), Tainan, Taiwan 70101

[4] National High Magnetic Field Laboratory, Tallahassee, FL, USA 32306

[5] School of Physics, Georgia Institute of Technology, Atlanta, GA, USA 30322

[6] Department of Physics, Florida State University, Tallahassee, FL, USA 32306

Weiwei Xie (weiweix@lsu.edu; weiwei.xie@rutgers.edu);
Martin Mourigal (mourigal@gatech.edu)

## *Abstract*

The localized *f*-electrons enrich the magnetic properties in rare-earth-based intermetallics. Among those, compounds with heavier 4*d* and 5*d* transition metals are even more fascinating because anomalous electronic properties may be induced by the hybridization of 4*f* and itinerant conduction electrons primarily from the *d* orbitals. Here, we describe the observation of trivalent $Yb^{3+}$ with $S = ½$ at low temperatures in $Yb_xPt_5P$, the first of a new family of materials. $Yb_xPt_5P$ ($0.20 \leq x \leq 1$) phases were synthesized and structurally characterized. They exhibit a large homogeneity width with the Yb ratio exclusively occupying the 1*a* site in the anti-$CeCoIn_5$ structure. Moreover, the resistivity measurement of a sample analyzed as $Yb_{0.25}Pt_5P$ shows it to exist a complete zero-resistance transition with a critical transition temperature of ~0.6 K, possible superconductivity. However, the zero-resistivity transition was not observed in $YbPt_5P$ with antiferromagnetic ordering existing solely. First-principles electronic structure calculations substantiate the antiferromagnetic ground state and indicate that 2D nesting around Fermi level may give rise to exotic physical properties, such as superconductivity. $Yb_xPt_5P$ appears to be a unique case among materials.

## *Introduction*

Understanding the electronic interactions in intermetallic compounds and designing the functional materials with targeted physical properties accordingly have been long-standing challenges in the materials science. One side, the classical chemical concepts, such as charge balance arguments and electron-counting rules, do not work for the intermetallic compounds with strong electron correlation, partially the valence orbital manifold, and occasionally relativistic effects.[1] On the other side, quantum-chemical techniques including machine-learning is hindered by the limited data of materials with specific properties, such as superconductivity.[2-4] The interplay between superconductivity and magnetism, which can happen under very restricted conditions, has the potential to lead to exotic new condensed matter physics and quantum devices. The coexistence of superconductivity and magnetism in a single material system is very rare.[5–10] Materials physicists have worked to realize this state by fabricating hybrid nanostructures that combine both superconducting and magnetic layers, and along similar lines chemists have used solvent methods to build up hybrid materials with superconducting and magnetic fragments, most of which are not fully ordered.[11–15] It is highly demanding to design and synthesize a new bulk material that displays the coexistence of superconductivity and magnetism in a single substance.

One chemical perspective for discovering new functional materials especially superconductors is to posit that similar physical properties can be observed in structural families. A well-known example is the $HoCoGa_5$-type structure motif type becoming intriguing after the discovery of heavy fermion superconductivity in $CeCoIn_5$[16]. The indium analogs, $CeTIn_5$ (T-Transition Metals) show an intricate interplay of superconductivity and magnetism, e.g. unconventional superconducting $CeCoIn_5$ and antiferromagnetic $CeRhIn_5$[17].

Heavy fermion superconductors, most of which are $4f^1$ Ce-based, are one way that the two kinds of electronic systems can interact,[16–19] and an alternative is for more weakly coupled rare earth-metal systems, such as is seen for rare earth Chevrel phases[6,20–23] and the lanthanide borocarbides.[24–28] $Yb^{3+}$, with a $4f^{13}$ electronic configuration, is often considered as the hole analogue of $Ce^{3+}$, however, only a single Yb-based heavy-fermion superconductor, $YbAlB_4$ has been reported to date, with $T_c$ = 0.08 K;[29] the large discrepancy must be due to unfavorable Yb-metal hybridization energies in most cases.

The ternary compound $LaPt_5As$, synthesized in rhombohedral symmetry under high pressure, provides a new avenue for research because, although non-magnetic, it hosts superconductivity with $T_c$~2.6 K.[30] Consisting of Pt-rich layered networks, superconducting $LaPt_5As$ inspired us to incorporate a magnetic rare earth element ($Yb^{3+}$) into the platinum-pnictide system. The much smaller ionic radius of Yb compared to La led us to replace $As^{3-}$ with smaller $P^{3-}$ to stabilize a hypothetical $YbPt_5P$ at ambient pressure. This may yield both superconducting and magnetic properties in a single material.

Thus, here we report a new material $Yb_xPt_5P$ with a thorough crystallographic and physical properties characterization. $Yb_xPt_5P$ crystallizes in tetragonal $TlPt_5As$-type structure with the space group *P*4/*mmm*. The structure can be considered as the anti-format of $CeCoIn_5$. According to single crystal X-ray diffraction, the Yb content (*x*) in $Yb_xPt_5P$ varies significantly from *x*= 0.20~1.0. We studied the magnetic and electronic properties on two samples with *x* = 0.23 (1) and 0.96 (1). In both samples, we observed antiferromagnetic transitions of $Yb^{3+}$ around 0.3 K. Moreover, the zero-resistivity transition, highly possible a superconducting transition, was observed around ~0.6 K only in low ratio Yb sample, $Yb_{0.20(1)}Pt_5P$, but not in $Yb_{0.96(1)}Pt_5P$. Our new quantum material is a new ideal platform to study the interplay between superconductivity and magnetism. Our new quantum material appears to be a distinct platform for studying the interactions between superconductivity and magnetism, in a material where the strong spin-orbit coupling is present.

## *Experimental Section*

**Synthesis:** $Yb_xPt_5P$ samples with several loading compositions ($x$= 0.4, 0.5, 0.8 and 1.1) were synthesized by a high-temperature solid-state method. Stoichiometric elemental Yb (<200 mesh, Alfa Aesar, ≥ 99.9%), Pt (~22 mesh, Beantown Chemical, ≥ 99.99%) and red P (~100 mesh, Beantown Chemical, ≥ 99%) were mixed well, and pressed into a pellet inside an argon-filled glovebox. The pellet was placed in an alumina crucible which was then sealed in an evacuated quartz tube. Heat treatment to 950 $^{o}$C was carried out with a rate of 30 $^{o}$C per hour in a Thermo Scientific furnace. After holding at 950 $^{o}$C for two days, the tubes were slowly cooled to room temperature in five days. Based on our experiments, heating at 950 $^{o}$C longer than 10 days would lead to the decomposition of $Yb_xPt_5P$ and a low ratio of Yb less than $x$ = 0.5 will less possibly yield the appropriate phases. Small single crystals (~0.4×0.2×0.02 $mm^3$) were attached to the bulk polycrystalline material, as shown in FIG. S1. In most of the cases, some impurities appeared as black powder, which can be removed by soaking in ethanol in an ultrasonic bath for 20 minutes. $Yb_xPt_5P$ is resistible to both air and moisture.

**Phase Identification:** The phase purity was determined by using a Rigaku MiniFlex 600 powder X-ray diffractometer (XRD) with Cu $K_\alpha$ radiation (λ=1.5406 Å, Ge monochromator). A long scan with the Bragg angle ranged from 3° to 90° in a step of 0.005° at a rate of 0.35°/min was performed for each sample. Rietveld method was utilized to fit the powder XRD pattern in the *Fullprof* Suite according to the calculated pattern from single crystal data.[31]

**Structure Determination:** Multiple pieces of crystals (~20×40×40 $\mu m^3$) were measured to get precise structural information. A Bruker Apex II diffractometer equipped with Mo radiation ($\lambda_{K\alpha}$= 0.71073 Å) was applied to explore the crystal structure at room temperature. The small crystals were stuck to a Kapton loop with glycerol. Four different positions were chosen to take the measurement with an exposure time of 10 seconds per frame and the scanning 2θ width of 0.5°. Direct methods and full-matrix least-squares on $F^2$ models with *SHELXTL* package were applied to solve the structure.[32] Data acquisition was obtained *via* Bruker *SMART* software with the corrections on Lorentz and polarization effect done by *SAINT* program. Numerical absorption corrections were accomplished with *XPREP*, which is based on the face-index modeling.[33]

**Physical Property Measurements:** All the physical property measurements were performed on pieces of as-grown samples extracted from the sample crucible. The measured pieces consisted of a mixture of polycrystalline matrix and single crystals which were semi-randomly oriented with respect to each other. Magnetization measurements were carried out for temperatures $T$ = 1.8 – 300 K using a Vibrating Sample Magnetometer (VSM) in Quantum Design PPMS systems. The heat capacity was measured for $T$ = 0.05 – 2 K using a Dilution Refrigerator (DR) and for $T$ = 2 – 200 K using the Heat Capacity (HC) option of the same Quantum Design PPMS systems. Electrical resistivity measurements were performed in a four-wire configuration with platinum or gold wires and silver contacts for $T$ = 0.1 – 300 K using the Adiabatic Demagnetization Refrigerator (ADR) option or using the combination of Dilution Refrigerator and Electrical Transport (ETO) options.

**Electronic Structure Calculation:** The bulk electronic structures of $YbPt_5P$ were computed using the projector augmented wave method[34,35] as implemented in the VASP package[36] within the generalized gradient approximation (GGA)[37] and GGA plus Hubbard $U$ (GGA+$U$)[38] scheme. On-site $U$ = 7 eV and 4 eV were used for Yb $f$-orbitals and Pt $d$-orbitals, respectively. The spin−orbit coupling (SOC) was included self-consistently in the calculations of electronic structures with a Monkhorst–Pack $k$-point mesh 20 × 20 × 10. The experimental structural parameters were employed.

**X-ray Photoelectron Spectroscopy (XPS):** The oxidation states of Yb, Pt and P atoms for $Yb_{0.67}Pt_5P$ and $YbPt_5P$ are determined by a Kratos AXIS 165 XPS/AES equipped with standard Mg/Al and high-performance Al monochromatic source in an evacuated ($10^{-9}$ torr) chamber at room temperature.

## *Results and Discussion*

**Phase Information, Crystal Structure and Chemical Composition Determination:** Single Crystal X-ray Diffraction (SXRD) analysis shows that $Yb_xPt_5P$ adopts the tetragonal structure illustrated in FIG. 1a, with space group *P*4/*mmm*, which can be considered an anti-$CeCoIn_5$-type[12] while in $Yb_xPt_5P$, Yb and Pt atoms are located on the 1*a* and 1*c* sites and in $CeCoIn_5$, the Ce and Co atoms occupy the 1*c* and 1*a* sites. The structure of $Yb_xPt_5P$ is layered, with planes of phosphorous atoms separating square-lattice layers of truncated $YbPt_{12}$ cuboids. These cuboids host two distinct Yb-Pt distances (i.e. Yb-Pt1: 2.876 (1) Å and Yb-Pt2: 2.810 (2) Å in $Yb_{0.96(1)}Pt_5P$.) The refined crystallographic data including atomic positions, site occupancies, and isotropic thermal displacements for the different Yb concentrations studied in detail are summarized in Tables S1 and S2 of the Supplementary Information. Our synthetic approach yielded $Yb_xPt_5P$ with various Yb ratios. By decreasing the occupancies of Yb on 1*a* site to ~23%, vacancies of P/Pt on 1*b*/4*i* sites appears. This aspect of the structural chemistry influences the bulk physical properties. Table 1 summarizes the synthetic results from the powder X-ray diffraction patterns and single crystal X-ray diffraction of selected samples. Of these, the low-Yb loadings yielded a mixture of 151-phase and unreacted Pt phase, which can be distinguished by the optical microscope. Attempts to stabilize the 151-phase with homogenous Yb occupancy by extending the annealing time results in the decomposition of 151-type phases. The powder X-ray diffraction patterns of $Yb_xPt_5P$ were shown in FIG.1b. It can be found that $Yb_xPt_5P$ phases were obtained with slight Pt or $PtP_2$ impurity with different occupancy of Yb. The samples used to perform the physical properties measurements were taken from the same specimen. After measurements done, the samples were ground into powder and the powder X-ray diffraction measurement confirmed the chemical compositions again.

**Table 1.** Compositions, phase analyses, lattice constants, and refined compositions for $Yb_xPt_5P$ phases. PXRD = powder X-ray diffraction; SCXRD = single crystal X-ray diffraction.

| Atomic % Yb Loaded | Impurities (Manual) | Phase (PXRD) | --- Composition --- (PXRD) | (SCXRD)[a] |
|---|---|---|---|---|
| 40 | Unreacted P; Pt | 151-type; Pt; P | $Yb_{0.23}Pt_5P_{0.90}$ | $Yb_{0.23(1)}Pt_{4.87(4)}P_{0.90(5)}$ |
| 80 | Pt; $PtP_2$ | 151-type; Pt; $PtP_2$ | $Yb_{0.67}Pt_5P$ | $Yb_{0.660(4)}Pt_5P$ |
| 110 | Pt | 151-type | $Yb_{0.96}Pt_5P$ | $Yb_{0.96(1)}Pt_5P$ |

[a] 296 K; Numbers in ( )'s are standard uncertainties.

For the X-ray powder diffraction patterns, all scale factors and lattice parameters were refined, while the displacement parameters of all atoms were assumed to be anisotropic. The refined lattice parameters for $Yb_xPt_5P$ phases showed a 1.54% and 2.64% increase along *a* and *c* according to X-ray powder diffraction as the Yb ratio increased from 25 to 100 atomic percent. Single crystals showed a similar trend. Analysis of samples all fall within various ratios of Yb in the phase. Does Yb still show 3+ oxidation state in the intermetallic $Yb_xPt_5P$? With the question in mind, X-ray Photoelectron Spectroscopy (XPS) experiments were performed on $Yb_{0.67}Pt_5P$ and $YbPt_5P$, which confirms 3+ oxidation state of Yb in both samples, as shown in FIG. S2.

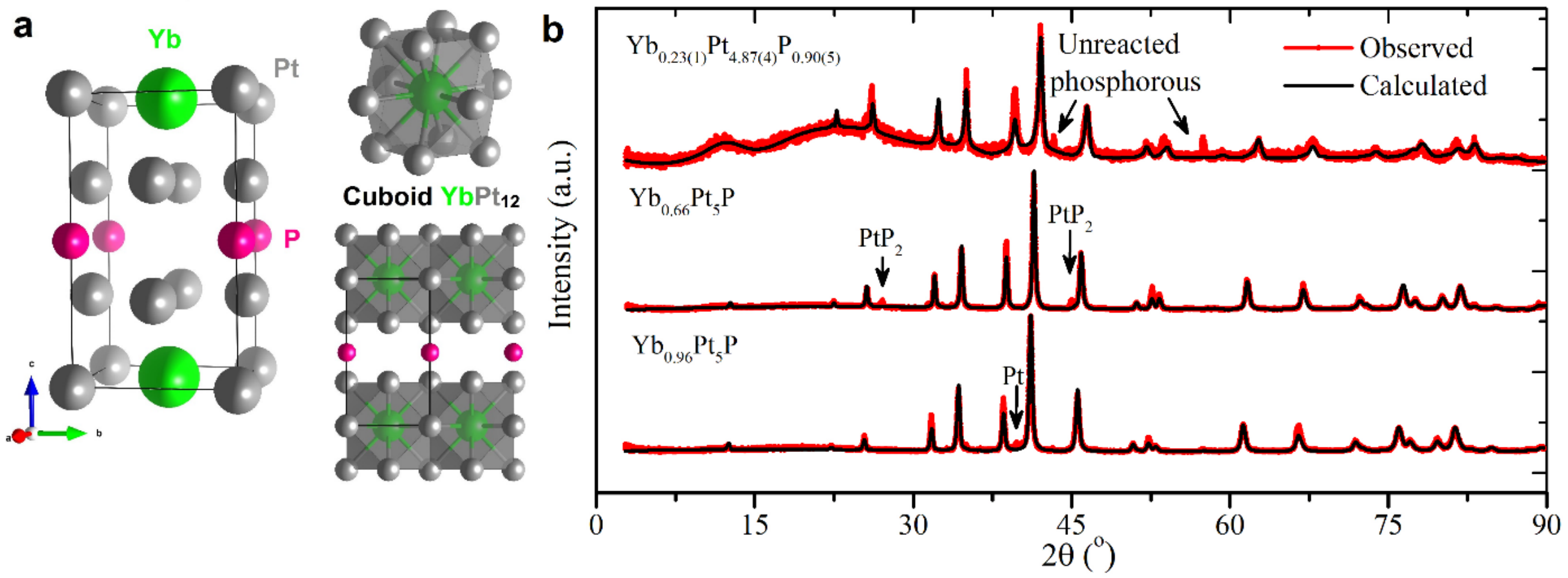

**FIG. 1 | Structural determination and phase characterization of $Yb_xPt_5P$. *a.*** The crystal structure of $YbPt_5P$, where green, grey, and red spheres represent Yb, Pt and P atoms, respectively. ***b.*** The Rietveld refinements of powder X-ray diffraction patterns of $Yb_xPt_5P$ with $x$ = 0.25, 0.67, and 1. The red line and dot indicate the observed reflection patterns and the black line represents the calculated pattern obtained from single crystal XRD. The calculated patterns and the peak positions of $YbPt_5P$ are indicated by green vertical ticks.

**Antiferromagnetic Ordering in $Yb_xPt_5P$:** The magnetic susceptibility is shown as the inset of FIG. 2*a* for low ratio Yb sample ($Yb_{0.25}Pt_5P$) with the chemical composition $Yb_{0.23(1)}Pt_{4.87(4)}P_{0.90(5)}$ confirmed by SEM-EDX. The data was fitted over the high-temperature region (HT from 225 to 300 K) and the low-temperature region (LT from 1.8 to 15 K) to the Curie-Weiss law without a diamagnetic correction. The effective moment of $\mu_{eff}$ = 4.21(5) $\mu_B$/f.u. obtained for the HT range is reduced to $\mu_{eff}$ = 1.82 (9) $\mu_B$/f.u. in the LT range, where it is associated with a negative Weiss temperature $\theta_W$ = -1.97 K indicative of an antiferromagnetic tendency. The isothermal magnetization at 2K in FIG. 2*a* (Main Panel) shows a negligible hysteresis and a saturation field around ~25 kOe with a saturation magnetization around 0.3 $\mu_B$/Yb. The heat capacity measurement, shown in FIG. 2*b*, illustrates a clearly magnetic transition peak around 0.23 K and the large entropy change in $Yb_{0.25}Pt_5P$.

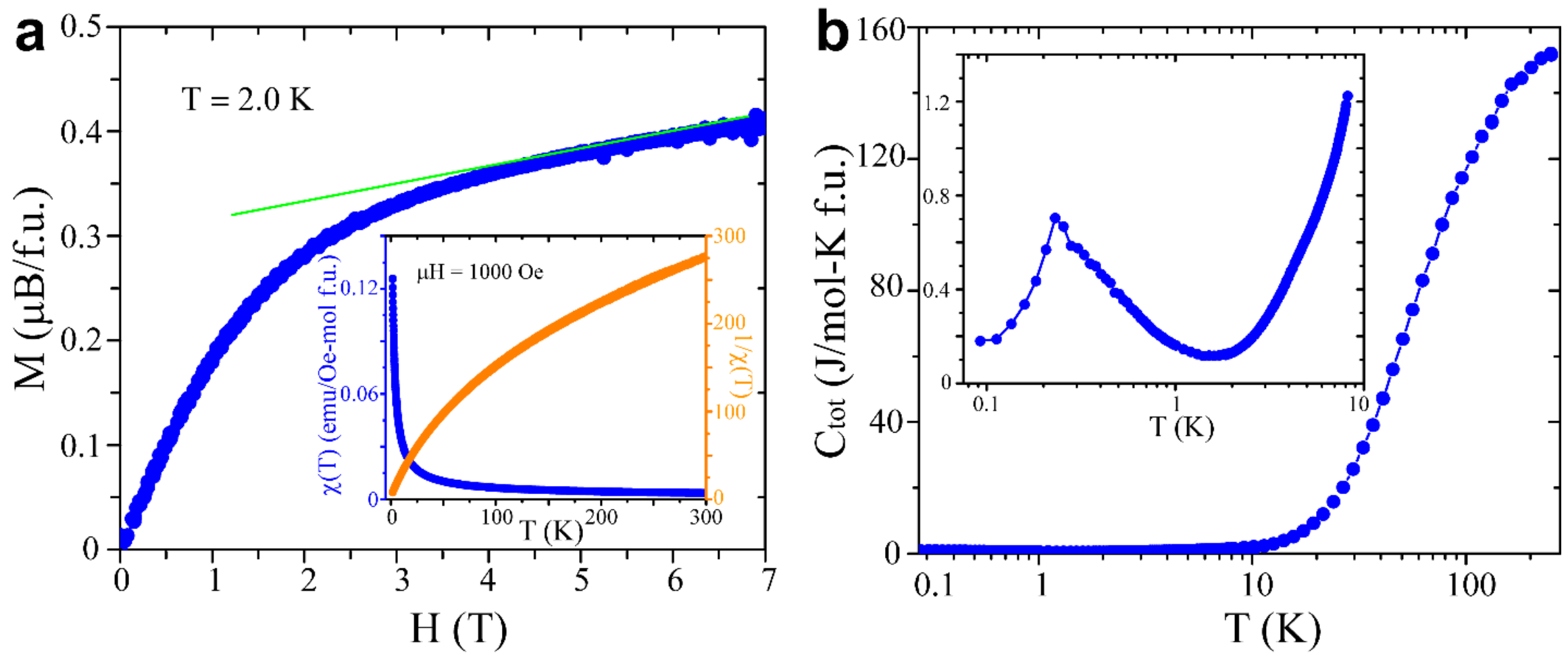

**FIG. 2 | Antiferromagnetism in $Yb_{0.25}Pt_5P$. *a.* (Main panel)** Isothermal magnetization up to 7 T at 2.0 K for $Yb_{0.25}Pt_5P$. **(Inset)** The temperature-dependence of the magnetic susceptibility for $Yb_{0.25}Pt_5P$ from 1.8 to 300 K measured under an applied field of 1000 Oe, data as indicated by the blue line for magnetic susceptibility and orange line for the inverse magnetic susceptibility. Green lines show the Curie-Weiss fitting of the inverse $\chi$ data at high temperature. ***b.*** (**Main panel**) Heat capacity measurements for $Yb_{0.25}Pt_5P$ from 0.1 to 150 K. **(Inset)** $C_{tot}/T$ vs $T^2$ for $Yb_{0.25}Pt_5P$ with magnetic transition occurring at 0.23 K.

As increasing Yb ratio to $YbPt_5P$, the magnetic characterization of $YbPt_5P$ is shown in FIG. 3*a* and *b*. The inverse susceptibility does not show any linear Curie-Weiss behavior until below 30 K, indicating low-lying crystal-electric-field (CEF) levels. These CEF levels is slightly lower than in other $Yb^{3+}$ systems in octahedral O environments. The Weiss constant is of the order of -2.2 to -2.7 K, which are strongly depending on the field directions and the fitting range. This indicates there is possibly never a real Curie-Weiss regime in this sample due to CEF levels (at high T range) and strong spin-orbit interactions between spins (at low T range). The field-dependent magnetization with different orientations in the magnetic field. Orientation "||" means the field along the long axis of the piece. Orientation "~" means the field perpendicular to the long axis of the piece. The results show only a weak orientation dependence with saturated magnetization values of Ms(||) = 2.7 $\mu_B$/f.u. and Ms(~) = 2.0 $\mu_B$/f.u. with the effective spin-1/2 and *g*-tensors of

5.4 and 4 depending on the field direction. The degree of spin-space anisotropy is consistent with observation of other $Yb^{3+}$ based insulating magnets. The normal heat capacity measurements up to 200 K in both 0 T and 10 T can be fitted using a double Debye model with an extraction of a "phonon background" for the extraction of entropy. The significant shift of the magnetic specific heat capacity between 0 T and 10 T is consistent with the large *g*-tensor. Without applied magnetic field, in addition to sharp peak at 0.28 K, there is a broad feature around 4 K that may be contributed from a low-lying CEF level. The magnetic entropy change (obtained after subtracting the "phonon background") indicates that there is indeed one effective *S*= 1/2 degrees of freedom per formula unit at temperatures below 3 K.

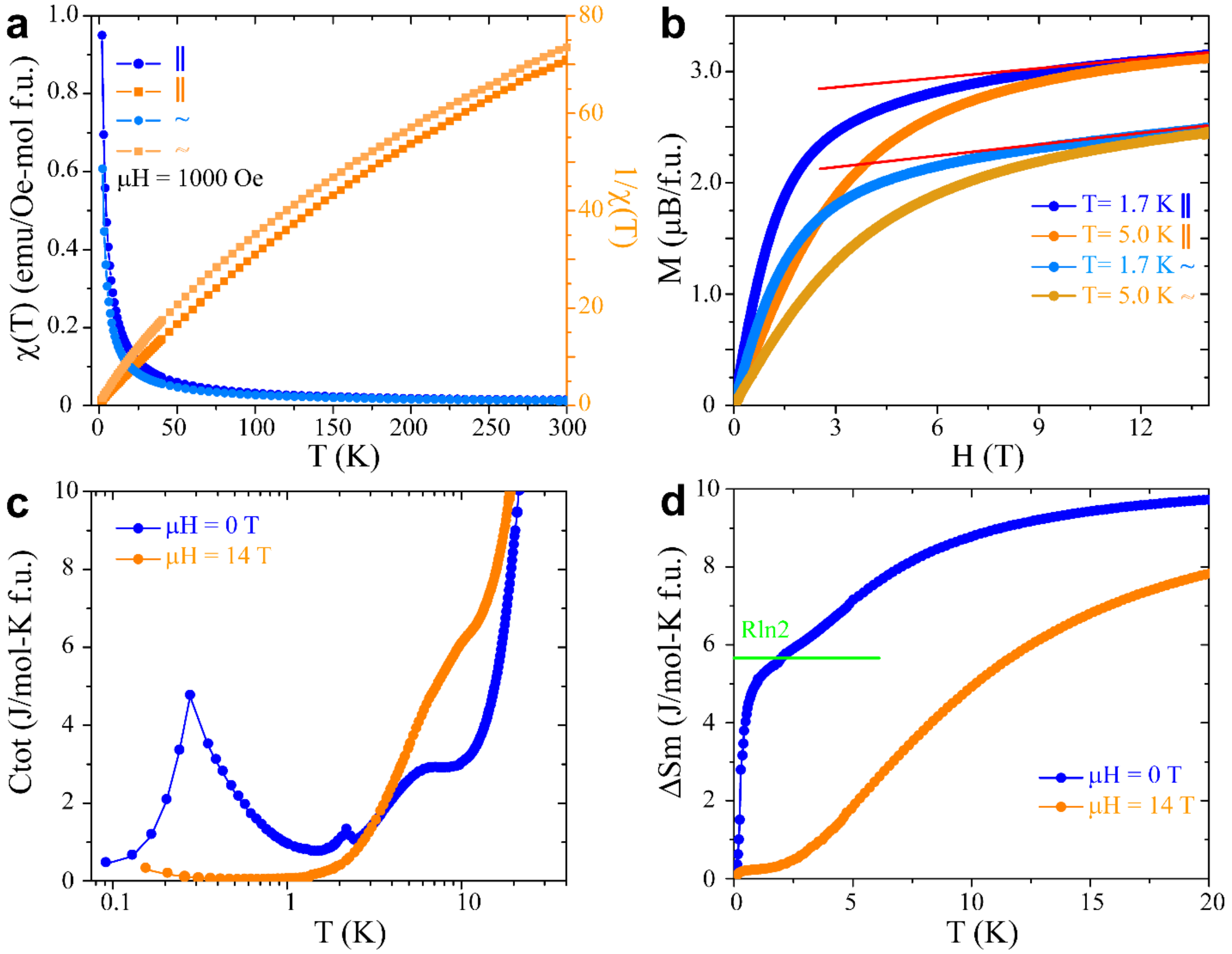

**FIG. 3 | Antiferromagnetism in $YbPt_5P$.** ***a.*** The temperature-dependence of the magnetic susceptibility for $YbPt_5P$ from 1.8 to 300 K measured under an applied field of 1000 Oe, data as indicated by the blue and light blue solid circles for the field along the long axis of the piece and the field perpendicular to the long axis of the piece, respectively. Orange and brown lines show the inverse temperature-dependence of the magnetic susceptibility. Green lines show the Curie-Weiss fitting of the inverse χ data at high temperature and low temperature, respectively. ***b.*** The magnetic behavior for $YbPt_5P$ up to 14 T at 1.7 and 5 K. ***c.*** Heat capacity measurement

with/without applied magnetic field with the emphasis on the magnetic ordering transition around 0.28 K. ***d.*** Entropy change related to the magnetic ordering of $Yb^{3+}$ in $YbPt_5P$ without applied field.

**Zero-resistance Transition Observed in $Yb_{0.25}Pt_5P$ but Suppressed in $YbPt_5P$:** FIG. 4*a* presents zero-resistance transitions of two samples with $x$ = 0.25 and 0.29. The resistivity curve from 1.8 to 300 K is consistent with what is expected for a metal in $Yb_{0.25}Pt_5P$ and $Yb_{0.29}Pt_5P$ without a phase transition shown in FIG. S3. The relatively small RRR may originate from defects on the Yb site. A drop to zero resistance is clearly seen at low temperatures, indicating the presence of a zero-resistance transition, highly possible a superconducting transition. The midpoints of the resistive transition of $Yb_{0.25}Pt_5P$ and $Yb_{0.29}Pt_5P$ are ~0.6 K and 0.65 K, respectively. To further characterize the zero-resistance transition, the inset of FIG. 2*a* shows the field-dependent resistivity curve of $Yb_{0.25}Pt_5P$. As applying magnetic field at various temperatures, the superconducting transition was suppressed gradually, which indicates that the strong spin-orbit coupling effects on Yb and Pt have negligible impact on the upper critical field of superconductivity. Moreover, the isostructural $Y_{0.34}Pt_5P$/$Y_{0.45}Pt_5P$ and $Pt_5P_2$ ($PtP_2$ is a semiconductor without resistance signal detected below 10 K.) were synthesized and characterized with no superconductivity observed above 0.4 K, as shown in FIG. S4 & FIG. S5 which can basically exclude the possibility that accidental impurities in Yb-based samples can contribute to the zero-resistance transition in resistivity measurements since both compounds were synthesized with identical procedure. The heat capacity measurements for $Yb_{0.29}Pt_5P$ in FIG. S6 shows a small kink around 0.6 K, which is consistent with the zero-resistance transition in resistivity. However, the large entropy changes from the magnetic transition of $Yb^{3+}$ in FIG. 2*b* make the subtle superconducting transition less possible to be observed. The similar problem occurs in the magnetic susceptibility measurements. Further study is required to confirm the superconductivity in the low ratio Yb samples. On the other side, the electric transport measurement of $YbPt_5P$ shows a failed superconducting transition starting from 0.6 K illustrated in FIG. 4*b*.

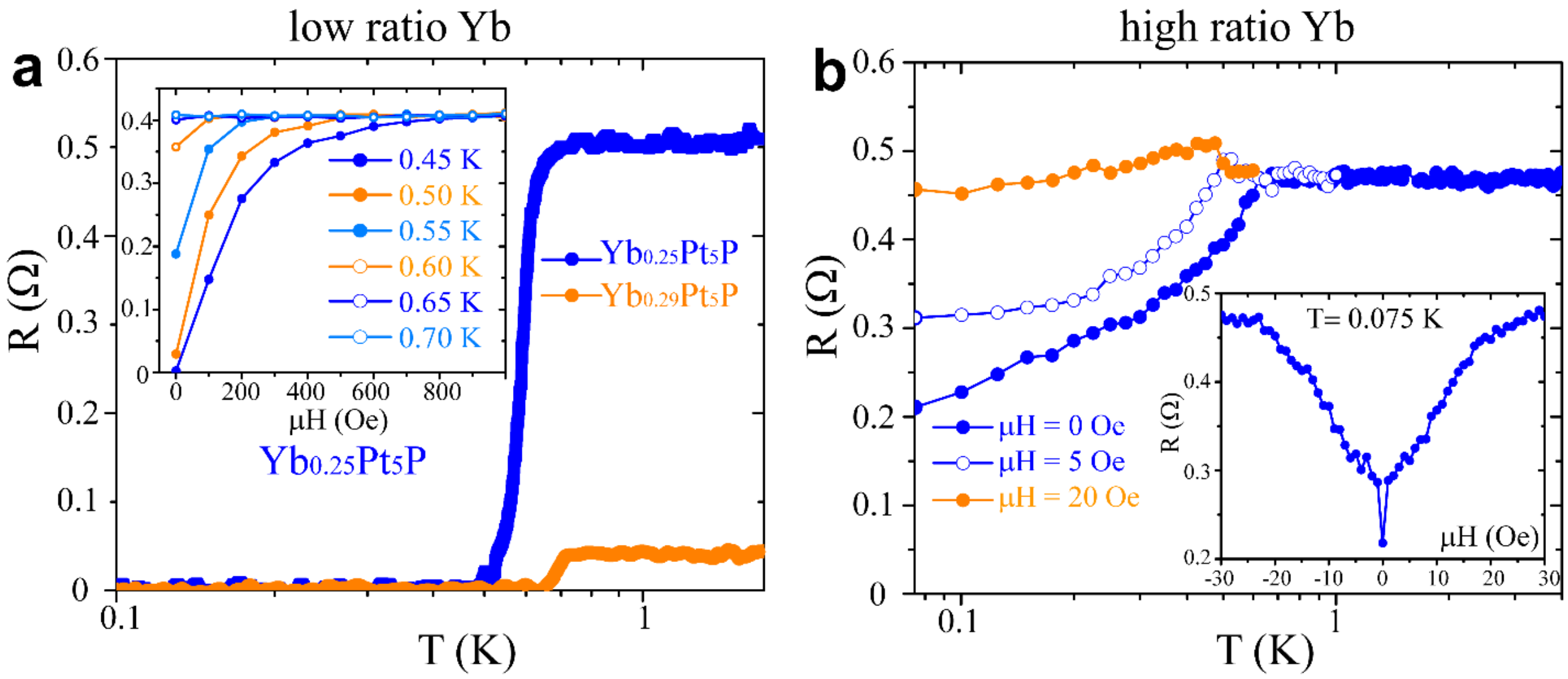

**FIG. 4 | *a.* Zero-resistance transition observed in low-Yb concentration samples for $Yb_{0.25}Pt_5P$ and $Yb_{0.29}Pt_5P$. (Main panel)** Electrical resistivity measurements for $Yb_{0.23(1)}Pt_{4.87(4)}P_{0.90(5)}$ and $Yb_{0.29(1)}Pt_5P$ from 0.1 to 300 K. **(Insert)** Field-dependent resistivity of $Yb_{0.23(1)}Pt_{4.87(4)}P_{0.90(5)}$ at various temperatures (0.45, 0.50, 0.55, 0.60, 0.65, 0.7 K). ***b.* Zero-resistance transition observed in low-Yb concentration samples** The temperature-dependent resistivity for $YbPt_5P$ from 0.05 to 4 K measured under various applied fields with ADR mode conducted with 100μA at 128Hz.

**Electronic Structure and 2D Nesting in $YbPt_5P$:** In order to further understand the nature of the ground state properties of $YbPt_5P$, we perform first-principles calculations on the bulk band structure based on generalized gradient approximation (GGA) and GGA with SOC (GGA+SOC) methods, as shown in FIG. 5. Our GGA and GGA+SOC calculations reveal a metallic ground state. From the orbital decomposition (FIG.5*a*), we find the flat and narrow Yb-4*f* bands are located around $E_F$ from -0.25 eV. Contrary to Yb-4*f* localized states, Pt-5*d* orbitals exhibit opposed behavior. The itinerant hole-like Pt-5*d* bands with a larger band dispersion span across $E_F$ and interact with the Yb-4*f* localized bands near $E_F$, leading to a complex Fermi surface (FS). The P-3*p* orbitals split into two components. Upper part displays electron-like band dispersion above $E_F$ and hybridizes strongly with Pt-5*d* bands, while lower part lies about −7 eV below $E_F$. FIG. 5*c* shows the three-dimensional (3D) FS of $YbPt_5P$, and the corresponding band numbers are labeled in FIG.5*d*. The FS of $YbPt_5P$ mainly contains four nearly 2D pockets (FIG.5*c* and *d*). Two tube-like and one bigger funnel-like hole-type pockets around Γ point (band-1, band-2, and band-3) and another one hole-type pocket around X point (band-3). These nearly 2D FS may induce a superconducting state or an antiferromagnetic magnetic (AFM) phase in $YbPt_5P$, resulted from the FS nesting effect. It is also note that the FS of $YbPt_5P$ is much different from the LDA band structure of heavy fermion $CeCoIn_5$ in which there are two Fermi sheets around M point and the FS display much stronger $k_z$ dispersion at Γ point. This difference implies the nature of ground state of $YbPt_5P$ may entirely distinct from the typical heavy fermion system. By projecting the band structure onto the cubic harmonics basis, we find the FS of the occupied bands around $E_F$ comes mainly from the Pt1-$d_{xy}/d_{z^2}$, Pt2-$d_{xy}$, and Yb-*f* (except Yb-$f_{xyz}$) orbitals (FIG. S7). The 3D real-space charge density distribution within the energy interval ($E_F$ ~ -10 meV) clearly shows this orbital anisotropy feature (FIG. S7). In this sense, the in-plane hopping strength is stronger than the out-of-plane one, consequently exhibition nearly 2D FS characteristic. When SOC is turned on (FIG.5*b*), Yb-4*f* bands split into the *j*=7/2 and *j*=5/2 states by SOC. The *j*=7/2 states dominate $E_F$ while *j*=5/2 states are shifted to -1.5 eV below $E_F$. In addition, strong SOC effect gap out the band crossing points between Pt-5*d* and P-3*p* orbitals around Γ point, and further enhance the band splitting of the Yb-4*f* and Pt-5*d* hybridized anti-crossing gap around $E_F$. Since the FS pattern of $YbPt_5P$ is significantly distinct from the heavy fermion Ce-115 family, the detailed theoretical modeling and experimental tests of the effect of *f*-*d* interaction in this material are left as an open question for future studies. Finally, we consider the electronic interactions via GGA plus correlation parameter U (GGA+U) calculations. We see the Yb-4*f* bands drop below $E_F$ as increasing the value of U (FIG. S9). As Yb-4*f* bands move to higher binding energies, their hybridization with conduction bands becomes smaller. The Pt-5*d* orbitals, on the other hand, are more pushed toward $E_F$ with increasing U and interact with P-3*p* states. Moreover, SOC effect split the Yb-4*f* state and further enhance the band splitting of crossing states, resulting in a continuous energy gap through whole Brillouin zone.

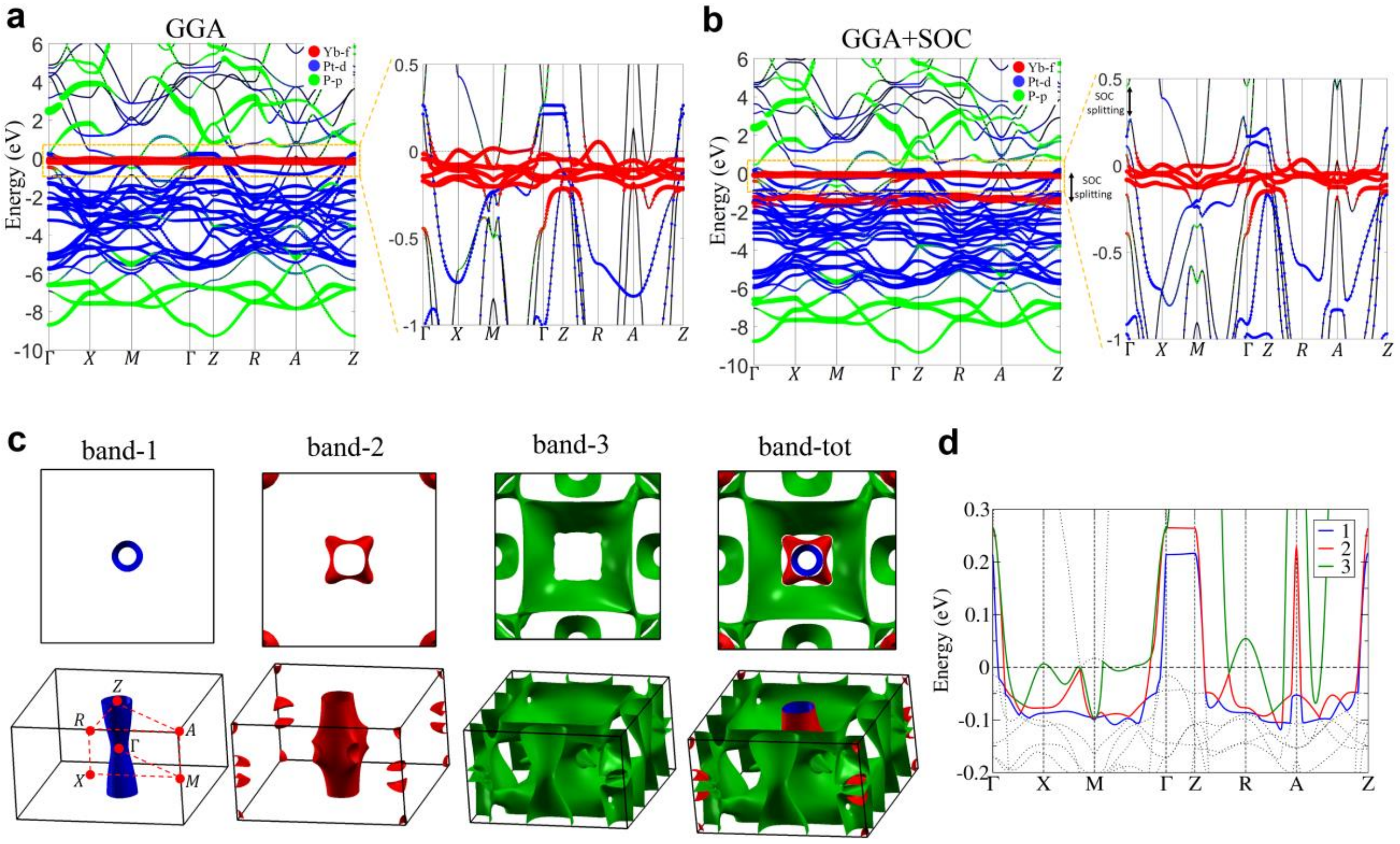

**FIG. 5 | Calculated bulk band structure and Fermi surface of $YbPt_5P$ in its non-magnetic phase.** ***a*.** The bulk band structure of $YbPt_5P$ based on GGA calculations without the inclusion of spin–orbit coupling. The Fermi energy is zero. The red, blue, and green dots indicate Yb-4*f*, Pt-5*d*, and P-3*p* orbitals, respectively. ***b*.** Same as **a** but with the inclusion of spin–orbit coupling (SOC). The Yb-4*f* bands split into the *j*=7/2 (around $E_F$) and *j*=5/2 (-1.5 eV below $E_F$) states by SOC. ***c*.** The bulk Fermi surface of non-magnetic $YbPt_5P$ based on GGA calculations. The corresponding band number are labeled in ***d***. Zoom-in band structure. The bands that cross Fermi level are labeled by blue, red, and green lines, respectively.

## Conclusions

$Yb_xPt_5P$ ($0.23 \leq x \leq 1$) phases exhibit superconducting trace and magnetism with low Yb-ratio and sole magnetism with high Yb-ratio. The isostructural material can be synthesized with other rare earths, with a variety of strange results expected. The coexistence of such strongly competing electronic states in a single substance itself makes this material remarkable. Moreover, the fact that the strong spin-orbit coupling of the Pt electrons must have an influence in determining the properties makes it truly novel. The complexity of the interactions between magnetism and superconductivity in this new materials family will push the frontiers of our knowledge of electronic and magnetic properties of materials into new areas and provide fertile ground for developing our still-emergent understanding of quantum materials. The material can be synthesized with other rare earths, with a variety of strange results expected.

## Supporting Information

Supplementary information, details of author contributions and competing interests; and statements of data are available at https://doi.org/xxx including:

Crystallographic data; Equivalent isotropic displacement parameters; Crystal picture; XPS data; Powder XRD pattern for $Yb_{0.29(1)}Pt_5P$; Resistivity measurement of $Y_{0.34}Pt_5P$, $Y_{0.45}Pt_5P$, and $Pt_5P_2$; GGA bulk band structures with projection of various orbitals onto the cubic harmonics basis; 3D real-space charge density distribution; bulk band structures of GGA+U and GGA+U+SOC.

## Acknowledgements

Weiwei Xie deeply appreciates Prof. Robert Cava's useful scientific discussion, generous support for testing superconductivity in various possible impurities down to 0.1 K in his lab at Princeton University, and continous encouragement. Weiwei Xie was also inspired by the interesting discussion with Profs. Dariusz Kaczorowski, Tomasz Klimczak, Shintaro Ishiwata.

The work at Xie's lab is supported by Beckman Young Investigator (BYI) Program and NSF-DMR-1944965. The work of M.D. and M.M. at Georgia Tech was supported by the National Science Foundation through Grant NSF- DMR-1750186. T.-R.C. was supported by the Young Scholar Fellowship Program from the Ministry of Science and Technology (MOST) in Taiwan, under a MOST grant for the Columbus Program MOST108-2636- M-006-002, National Cheng Kung University, Taiwan, and National Center for Theoretical Sciences, Taiwan. This work was supported partially by the MOST, Taiwan, Grant MOST107-2627-E-006-001. A portion of this work was performed at the National High Magnetic Field Laboratory, which is supported by the National Science Foundation Cooperative Agreement No. DMR-1644779 and the State of Florida. K.W. acknowledges the support of the Jack E. Crow Postdoctoral Fellowship. Electrical transport, magnetization, and heat capacity measurements performed by REB were supported by the Center for Actinide Science and Technology, an Energy Frontier Research Center funded by the U.S. Department of Energy (DOE), Office of Science, Basic Energy Sciences (BES), under Award No. DESC0016568.

## Supplementary Information

# A Novel Magnetic Material by Design: Observation of $Yb^{3+}$ with Spin-1/2 and Possible Superconducting Trace in $Yb_xPt_5P$

*Xin Gui,*[1] *Tay-Rong Chang*[2,3] *Kaya Wei,*[4] *Marcus J. Daum,*[5] *David E. Graf,*[4] *Ryan. E. Baumbach,*[4,6] *Martin Mourigal,*[5*] *Weiwei Xie*[1*]

[1] Department of Chemistry, Louisiana State University, Baton Rouge, LA, USA 70803

[2] Department of Physics, National Cheng Kung University, Tainan, Taiwan 70101

[3] Center for Quantum Frontiers of Research & Technology (QFort), Tainan, Taiwan 70101

[4] National High Magnetic Field Laboratory, Tallahassee, FL, USA 32306

[5] School of Physics, Georgia Institute of Technology, Atlanta, GA, USA 30322

[6] Department of Physics, Florida State University, Tallahassee, FL, USA 32306

Weiwei Xie (weiweix@lsu.edu; weiwei.xie@rutgers.edu);

Martin Mourigal (mourigal@gatech.edu)

## Table of Contents

**Table S1.** Single crystal structure refinement for $Yb_xPt_5P$ at 296 (2) K.

| Refined Formula | $Yb_{0.23(1)}Pt_{4.87(4)}P_{0.90(5)}$ | $Yb_{0.29(1)}Pt_5P$ | $Yb_{0.663(4)}Pt_5P$ | $Yb_{0.96(1)}Pt_5P$ |
|---|---|---|---|---|
| F.W. (g/mol) | 1017.76 | 1056.60 | 1120.63 | 1172.54 |
| Space group; Z | *P* 4/*mmm*; 1 | *P* 4/*mmm*; 1 | *P* 4/*mmm*; 1 | *P* 4/*mmm*; 1 |
| *a*(Å) | 3.9001 (3) | 3.911 (1) | 3.951 (1) | 3.974 (2) |
| *c*(Å) | 6.8108 (5) | 6.842 (2) | 6.953 (2) | 7.009 (4) |
| *c*/*a* | 1.746 | 1.749 | 1.760 | 1.764 |
| V (Å$^3$) | 103.60 (2) | 104.68 (7) | 108.56 (6) | 110.7 (1) |
| Extinction Coefficient | 0.026 (3) | 0.012 (2) | 0.0076 (8) | 0.0028 (3) |
| θ range (°) | 2.991-33.141 | 2.977-33.039 | 2.930-33.314 | 2.906-33.025 |
| No. reflections; $R_{int}$ | 1848; 0.0521 | 1130; 0.0514 | 2878; 0.0418 | 2589; 0.0372 |
| No. independent reflections | 152 | 153 | 163 | 164 |
| No. parameters | 15 | 13 | 13 | 13 |
| $R_1$: $\omega R_2$ ($I$>2δ($I$)) | 0.0311; 0.0762 | 0.0309; 0.0784 | 0.0189; 0.0499 | 0.0176; 0.0319 |
| Goodness of fit | 1.438 | 1.072 | 1.349 | 1.108 |
| Diffraction peak and hole (e$^-$/ Å$^3$) | 4.757; -6.559 | 4.037; -4.968 | 2.335; -2.759 | 1.993; -3.909 |

**Table S2.** Atomic coordinates and equivalent isotropic displacement parameters for $Yb_xPt_5P$ at 296 (2) K. ($U_{eq}$ is defined as one-third of the trace of the orthogonalized $U_{ij}$ tensor (Å$^2$))

| Formula | Atom | Wyckoff. | Occ. | *x* | *y* | *z* | $U_{eq}$ |
|---|---|---|---|---|---|---|---|
| $Yb_{0.23(1)}Pt_{4.87(4)}P_{0.90(5)}$ | Pt1 | 1*c* | 1 | 0 | ½ | 0.2875 (1) | 0.0029 (4) |
| | Pt2 | 4*i* | 1 | 0 | 0 | 0 | 0.0046 (4) |
| | Yb3 | 1*a* | 0.23 (1) | ½ | ½ | 0 | 0.017 (3) |
| | P4 | 1*d* | 0.90 (5) | 0 | 0 | ½ | 0.001 (2) |
| $Yb_{0.29(1)}Pt_5P$ | Pt1 | 1*c* | 1 | 0 | ½ | 0.2892 (1) | 0.0069 (4) |
| | Pt2 | 4*i* | 1 | 0 | 0 | 0 | 0.0084 (5) |
| | Yb3 | 1*a* | 0.29 (1) | ½ | ½ | 0 | 0.010 (2) |
| | P4 | 1*d* | 1 | 0 | 0 | ½ | 0.007 (2) |
| $Yb_{0.663(4)}Pt_5P$ | Pt1 | 1*c* | 1 | 0 | ½ | 0.2943 (1) | 0.0070 (2) |
| | Pt2 | 4*i* | 1 | 0 | 0 | 0 | 0.0086 (2) |
| | Yb3 | 1*a* | 0.663 (4) | ½ | ½ | 0 | 0.0075 (4) |
| | P4 | 1*d* | 1 | 0 | 0 | ½ | 0.0068 (7) |
| $Yb_{0.96(1)}Pt_5P$ | Pt1 | 1*c* | 1 | 0 | ½ | 0.2966 (1) | 0.0048 (2) |
| | Pt2 | 4*i* | 1 | 0 | 0 | 0 | 0.0052 (3) |
| | Yb3 | 1*a* | 0.96 (1) | ½ | ½ | 0 | 0.0042 (5) |
| | P4 | 1*d* | 1 | 0 | 0 | ½ | 0.008 (2) |

**FIG. S1 |**The crystal attached to a polycrystalline matrix.

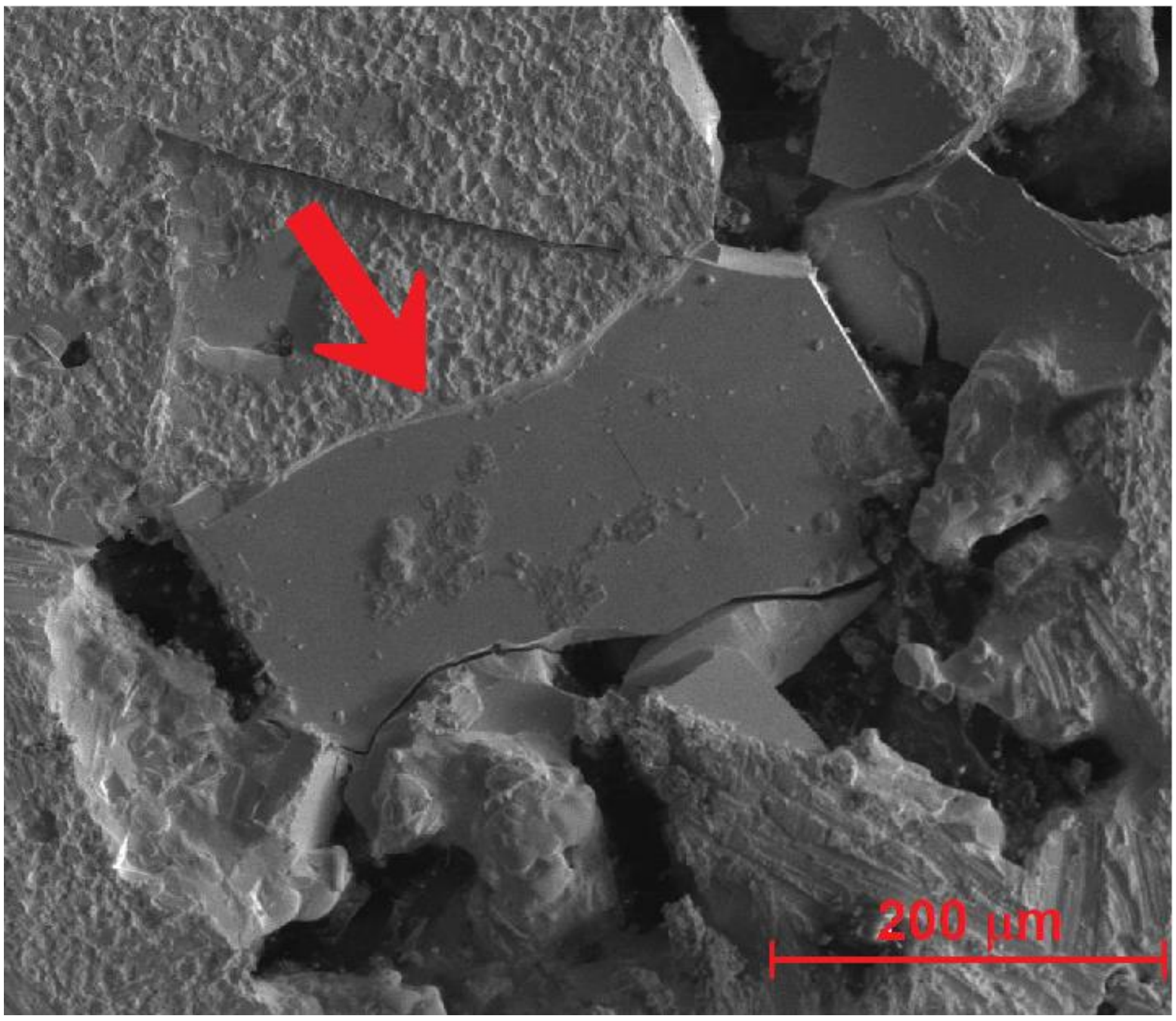

**FIG. S2.** XPS results of **(A).** Yb & P in $Yb_{0.663(4)}Pt_5P$; **(B).** Pt in $Yb_{0.663(4)}Pt_5P$; **(C).** Yb & P in $Yb_{0.96(1)}Pt_5P$; **(D).** Pt in $Yb_{0.96(1)}Pt_5P$.

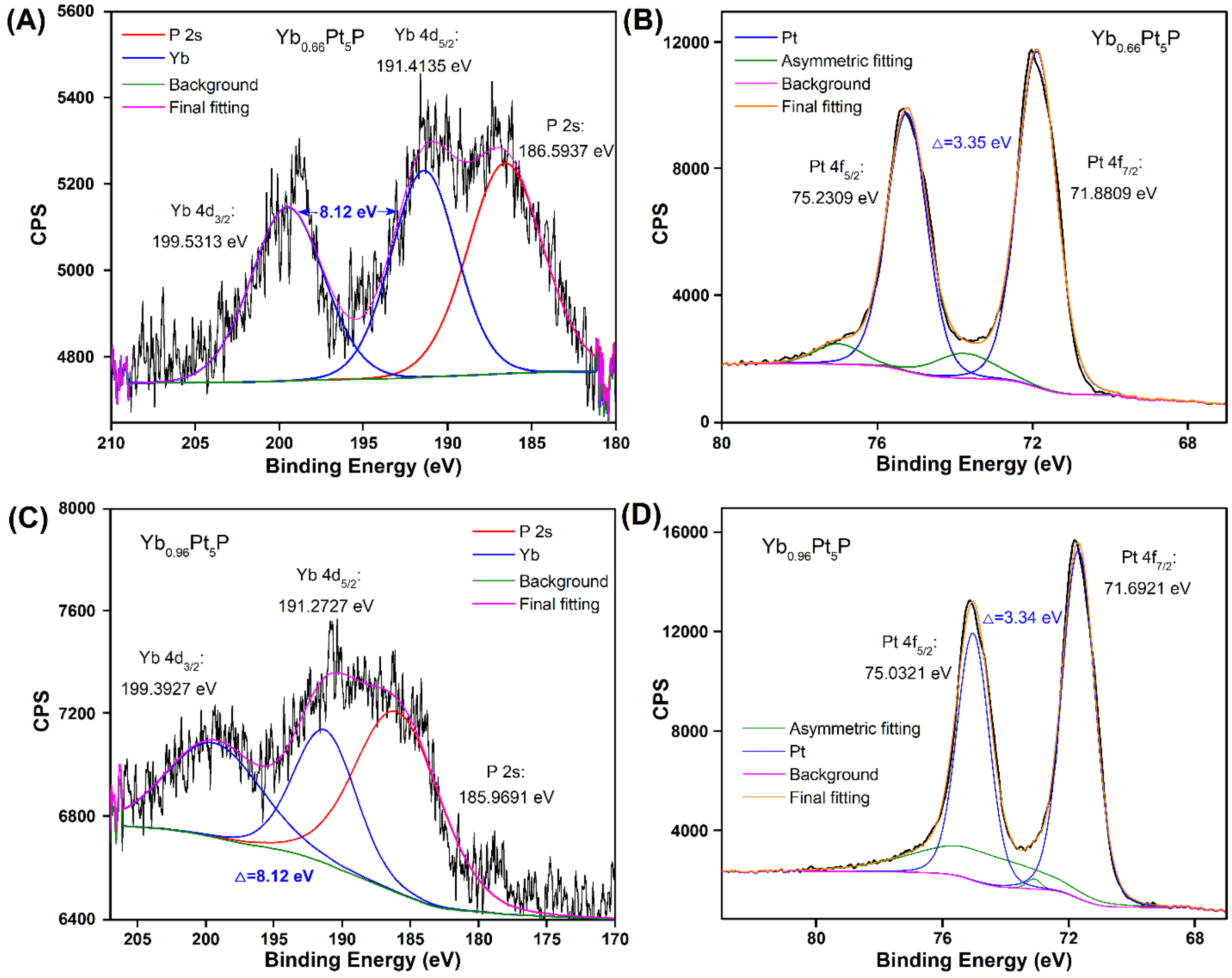

**FIG. S3.** Observed and calculated powder X-ray diffraction pattern for $Yb_{0.29(1)}Pt_5P$. Blue and pink lines represent observed and calculated patterns and orange arrows stand for impurities peaks.

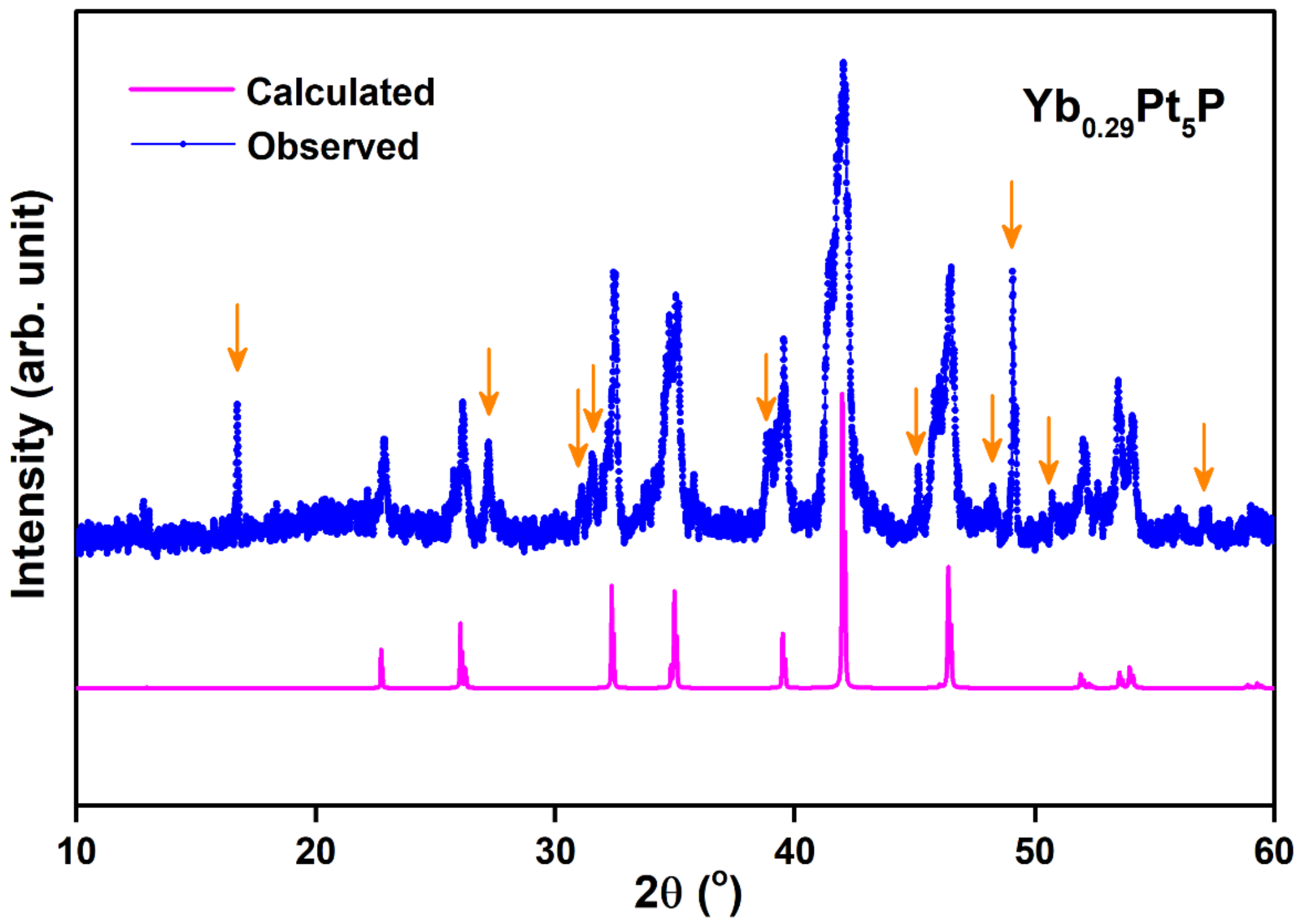

**FIG. S4.** The resistivity measurement of $Y_{0.34}Pt_5P$ and $Y_{0.45}Pt_5P$.

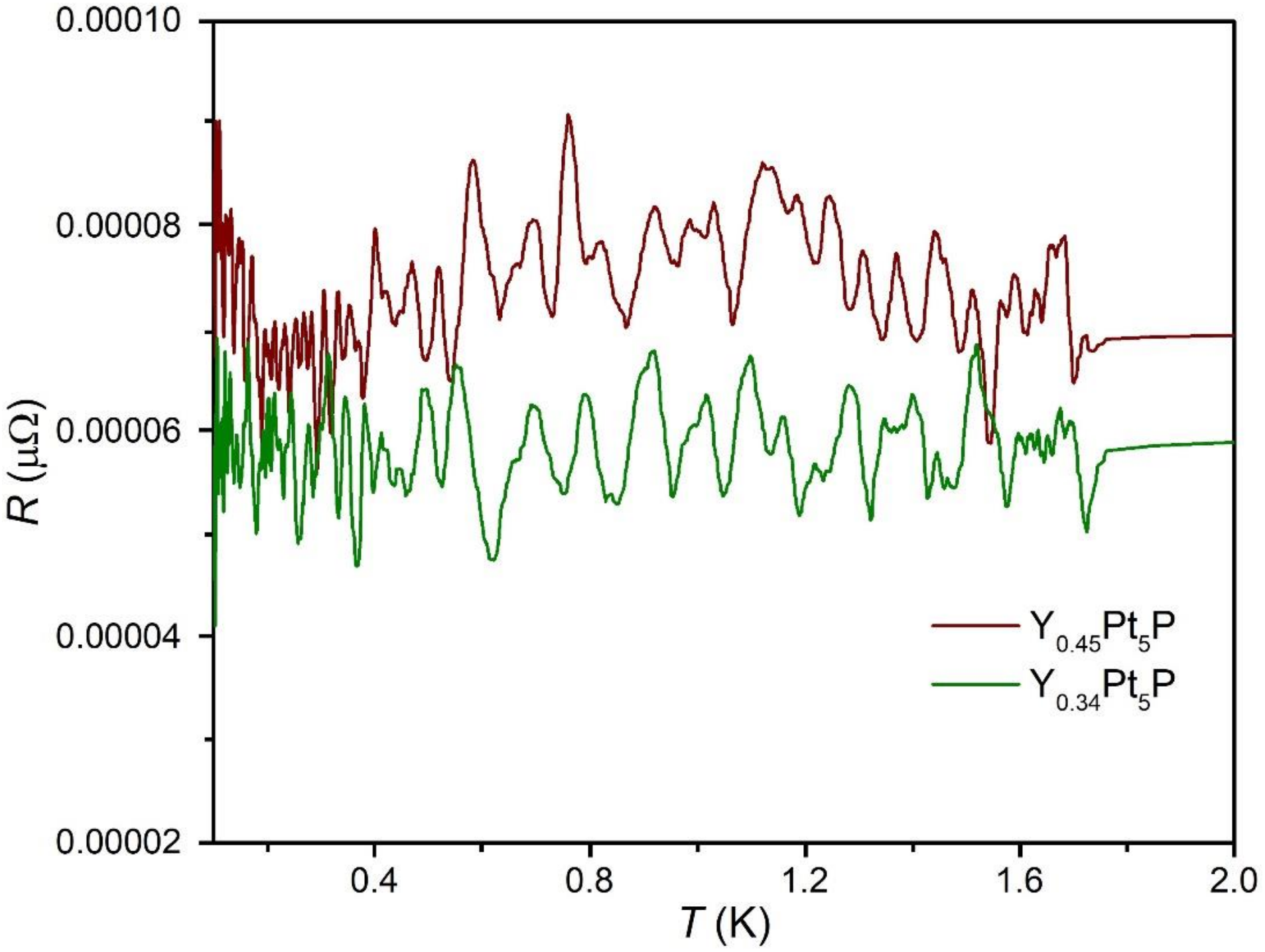

**FIG. S5.** The resistivity measurement of $Pt_5P_2$.

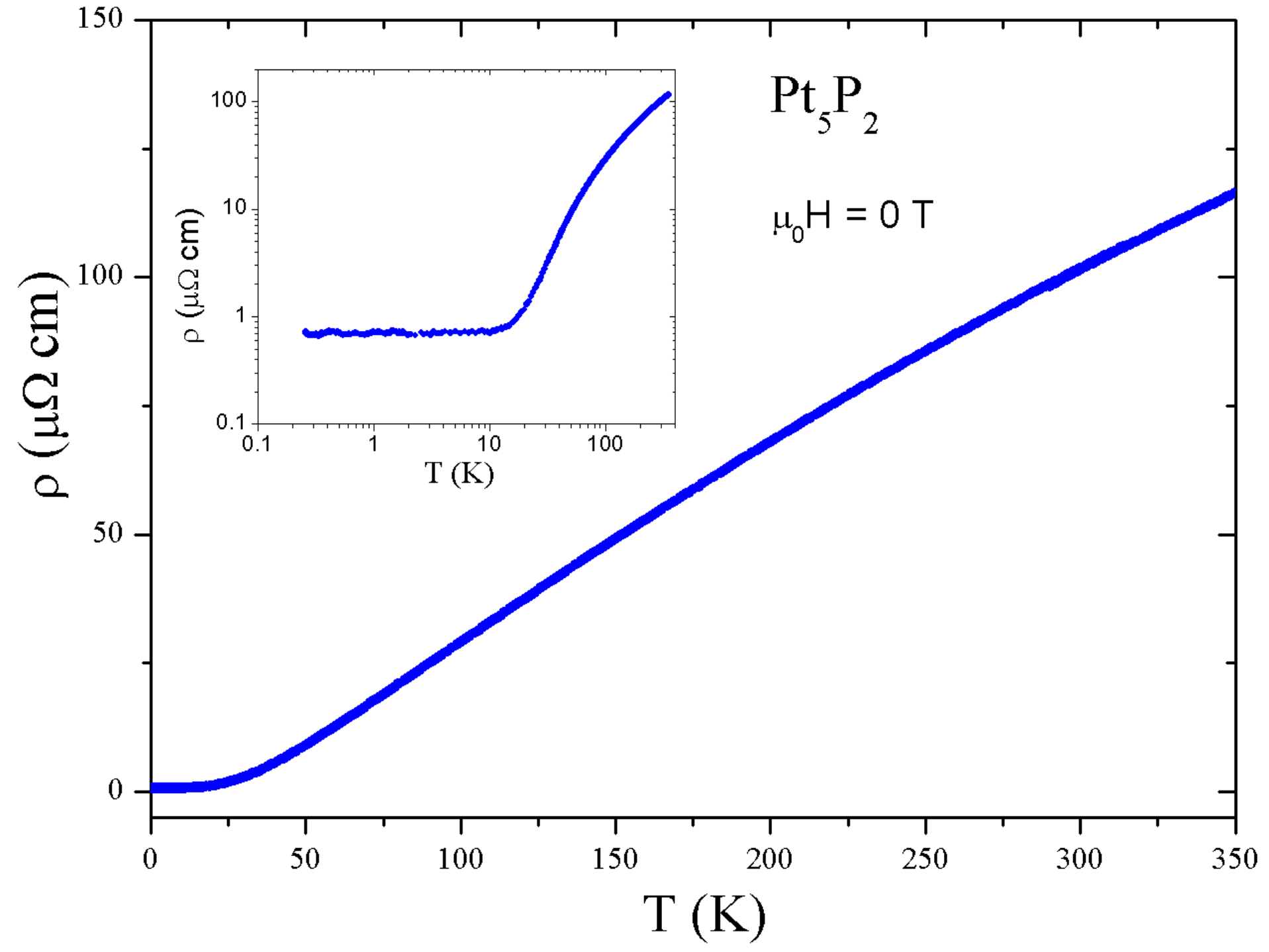

**FIG. S6.** The GGA bulk band structures zoom-in around $E_F$ but project onto the cubic harmonics basis.

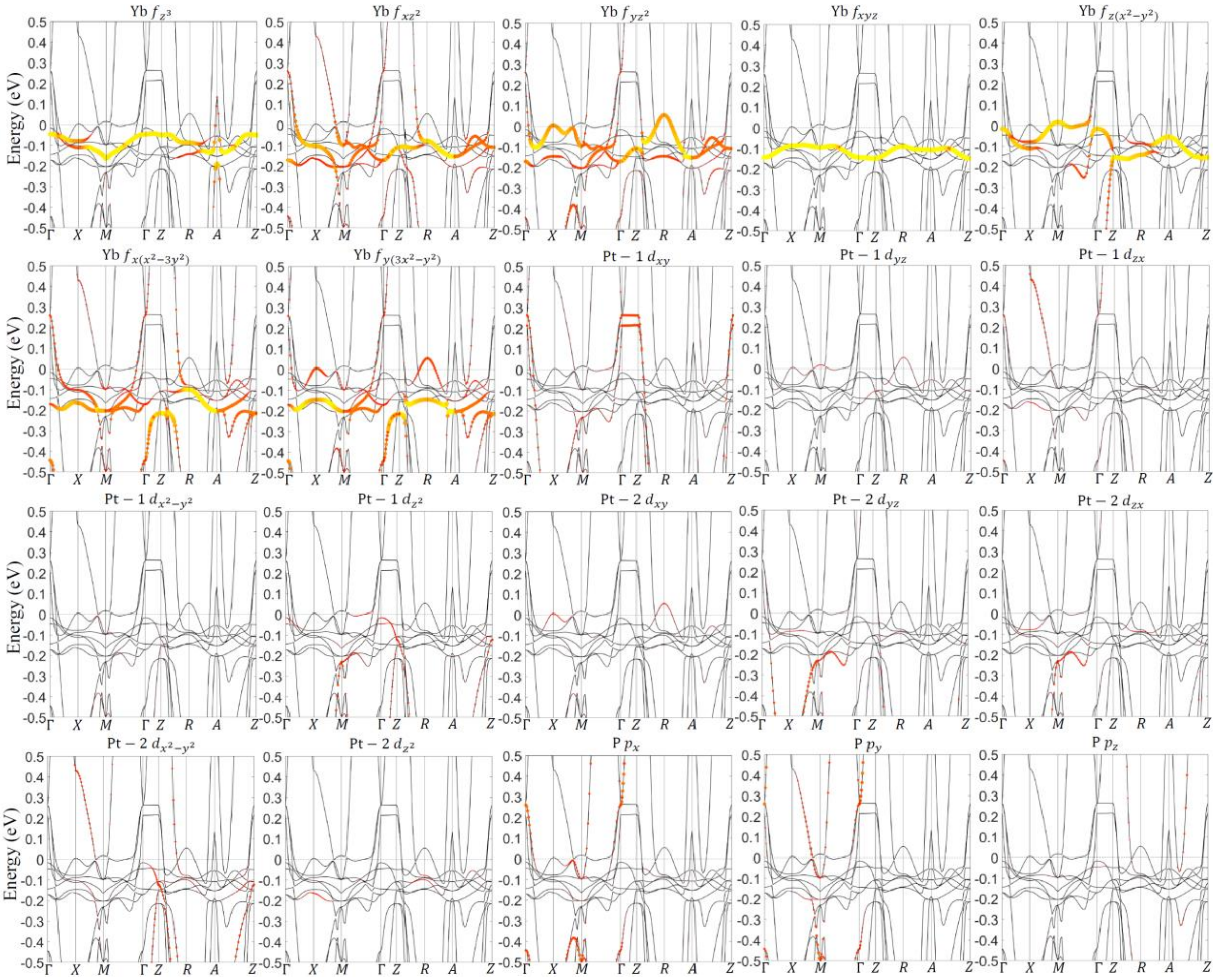

**FIG. S7.** Three-dimensional (3D) real-space charge density distribution within the energy interval ($E_F$ ~ -10 meV). The orange, blue, and red balls indicate Yb, Pt, and P atoms, respectively. The occupied orbitals on the Pt1 sites and Pt2 sites are mainly from $d_{xy}/d_{z^2}$ and $d_{xy}$, respectively. In contrast, the Yb-4*f* states show the more isotropic distribution. Since the charge density lobe of Pt2-$d_{xy}$ is directly point to the Yb-4*f* states, the in-plane hopping strength is stronger than the out-of-plane one that manly form interlayer Pt1- $d_{z^2}$ hopping.

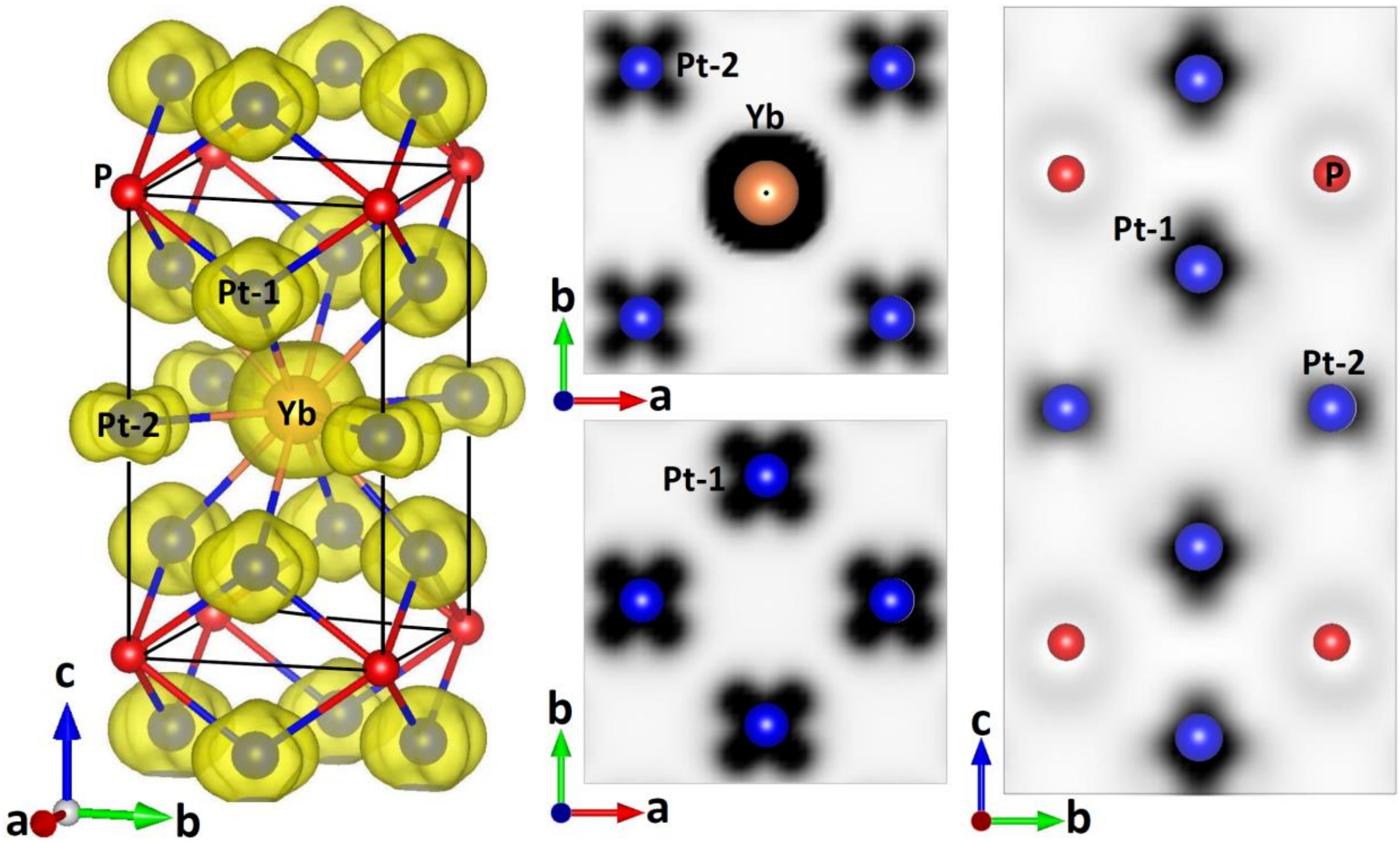

**FIG. S8.** The bulk band structures of GGA+U and GGA+U+SOC.

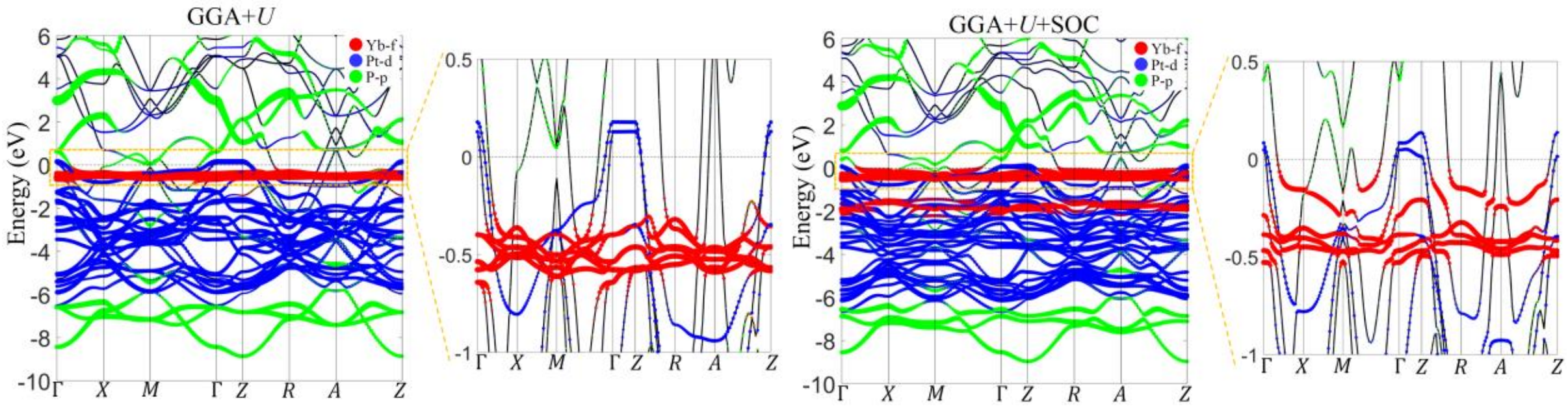